\documentclass{iaus}
\usepackage{graphicx}

\def\msun{\ifmmode {\rm M}_{\mathord\odot}\else $M_{\mathord\odot}$\fi}
\newcommand{\apj} {{\it ApJ}}

\newcommand{\avg}[1]  {{\langle #1 \rangle}} 
\newcommand{\msd} {M_{\rm *d}}

\newcommand{\mdin} {\dot M_{\rm in}}
\newcommand{\csd} { c_{\rm s,d}}
\newcommand{\omkin} { \Omega_{\rm k, in}}

\title[The Effects of Radiation Feedback on Early Fragmentation and Multiplicity] 
{The Effects of Radiation Feedback on Early Fragmentation and
  Stellar Multiplicity}

\author[Stella S. R. Offner]   
{Stella S. R. Offner$^1$
 }

\affiliation{$^1$Harvard-Smithsonian Center for Astrophysics, 60 Garden St, Cambridge MA 02138, USA \\ email: {\tt soffner@cfa.harvard.edu}} 
\pubyear{2010}
\volume{IAUS270}  
\pagerange{??--??}
\jname{Computational Star Formation}
\editors{J. Alves, B. Elmegreen, J. Girart \& V. Trimble, eds.}
\begin{document}

\maketitle

\begin{abstract}
Forming stars emit a significant amount of radiation into their natal
environment. While the importance of radiation feedback from high-mass
stars is widely accepted, radiation has generally been ignored in simulations
of low-mass star formation. I use ORION, an adaptive mesh refinement (AMR)
three-dimensional gravito-radiation-hydrodynamics code, to model
low-mass star formation in a turbulent molecular cloud. 
I demonstrate that including
radiation feedback has a profound effect on fragmentation and
protostellar multiplicity. 
Although heating is mainly confined within the core
envelope, it is sufficient to suppress disk
fragmentation that would otherwise result in low-mass companions
or brown dwarfs. As a consequence, turbulent fragmentation, not disk fragmentation, is likely the origin of
low-mass binaries. 

\end{abstract}

\firstsection
\section{Introduction}

The origin of stellar multiplicity remains an unsolved problem
in star formation. The dense conditions in molecular cloud cores and
dim luminosities of young protostars make estimations of the initial
multiplicity distribution challenging (\cite[Duch{\^e}ne
et~al. 2007]{duchene07}).  
However, the present-day multiplicity can be observed among field
stars, where the likelihood of companions
is strongly correlated with the primary stellar
mass. Nearly all O and B stars are found in binaries or multiple
systems, while only $\sim$20\% of M stars have companions
(\cite[Lada 2006]{lada06}). 
Successful simulations and theories of star formation must be able to
predict the multiplicity fraction and explain why it depends so strongly on stellar mass.
 
A number of mechanisms have been proposed of which two appear to have
the most potential for producing the observed number of multiple
systems (see \cite[Tohline 2002]{tohline02} for a review). 
First, gravitational instability within a protostellar accretion disk may produce
companions within a few 100 AU (\cite[Adams et
  al. 1989]{adams89}; \cite[Bonnell et al. 1994]{bonnell94}). 
Repeated fragmentation over the disk lifetime may generate numerous companions.
Second, perturbations within a turbulent core may seed additional
fragmentation on scales of $\sim$ 0.001-0.1 pc  (\cite[Fisher
  2004]{fisher04}; \cite[Goodwin et al. 2004]{goodwin04}). This must occur within the first $\sim$ 0.5 Myr of
collapse, when the core still contains at least $0.1 \msun$. 

In this paper, we discuss the effect of radiation feedback on early
fragmentation and stellar multiplicity using 3D adaptive mesh
refinement (AMR) simulations of
turbulent molecular clouds. We compare the cases with and without
radiation from forming stars. We describe the simulations
in \S2, present results in \S3, and conclude in \S4. 

\section{Simulations}

The ORION code solves the equations of compressible gas
dynamics, Poisson equation, and radiation energy equation in the flux-limited
diffusion approximation (\cite[Krumholz et al. 2007]{krumholz07}):
\begin{eqnarray}
{{\partial \rho} \over{ \partial t}} + \nabla \cdot (\rho {\bf v}) &=& 0, \\
{{\partial (\rho {\bf v})} \over{ \partial t}} + \nabla \cdot (\rho {\bf vv}) &=& -\nabla P - \rho \nabla \phi, \\
{{\partial (\rho e)} \over{ \partial t}} + \nabla \cdot [(\rho e + P){\bf v}] &=& \rho {\bf v} \nabla \phi - \kappa_R \rho (4\pi B -cE), \\
{{\partial E} \over{ \partial t}} - \nabla \cdot (\frac{c \lambda }{ \kappa_{\rm R} \rho} \nabla E) &=& \kappa_{\rm P} \rho (4 \pi B - cE) + \sum_n L_n \delta({\bf x}-{\bf x}_n) \label{radenergy}, \\
{\nabla}^2 \phi &=& 4 \pi G  [ \rho + \sum_n m_n \delta({\bf x}-{\bf x}_n) ],
\end{eqnarray}
where $\rho$, $P$, $\bf v$ and $e$ are the fluid density, pressure,
velocity, and specific kinetic plus internal energy of the gas, $\phi$ is the  gravitational
potential, $m_n$, ${\bf
  x}_n$, and $L_n$ are the mass, position, and luminosity of the
n$^{\rm th}$ star, $E$ is the radiation energy density, and
$\kappa_{\rm R}$ and $\kappa_{\rm P}$ are the Rosseland and Planck
dust opacities. 
For comparison, a second calculation closes the equations with a
barotropic equation of state (EOS) in lieu of equation \ref{radenergy}:
\begin{equation}
P = \rho c_{\rm s} ^ 2 +  \left({{\rho} \over {\rho_{\rm c}}}\right)^ {\gamma} \rho_{\rm c} c_{\rm s}^2, 
\end{equation}
where  $c_{\rm s} = ({k_{\rm B} T }/{ \mu})^{1/2}$ is the sound speed,
$\gamma=5/3$, the average molecular weight $\mu =2.33m_{\rm H}$, and
the critical density, $\rho_{\rm c}=2\times 10^{-13}$ g cm$^{-3}$. 

The calculations insert Lagrangian sink particles in regions
exceeding the Jeans density on the maximum AMR level 
(\cite[Krumholz et al. 2004]{krumholz04}). In the radiation
simulation, particles have a sub-grid model based upon \cite[McKee \& Tan
  (2003)]{mckee03} that includes the
accretion energy, Kelvin-Helmholtz
contraction, and nuclear burning (\cite[Offner et al. 2009]{Offner09}).

The calculations have a Mach number of  ${{\mathcal
    M}_{\rm 3D}}$=6.6, domain size $L=0.65$ pc,  and mass $M$=185 \msun, which
correspond to an approximately virialized cloud. We adopt
periodic boundary conditions for the gas and Marshak boundary
conditions for the radiation field, which allows the cloud to cool. We use a $256^3$ base grid with 4 levels of grid refinement, where $\Delta x_4$  =  32 AU. High-resolution
    convergence tests are discussed in detail in \cite[Offner et al. (2009)]{Offner09}.
We drive the boxes for three crossing times using random
velocity perturbations with wavenumbers $1 \le k \le 2$ after which self-gravity is turned on. The
initial gas temperature is 10 K. Since the gas cools efficiently
during the driving phase, the radiation calculation remains
nearly isothermal.

\section{Results}

\subsection{Fragmentation}

At the end of a freefall time, the two calculations have very
different temperature distributions. Temperatures in the EOS calculation do not exceed 15 K, while temperatures in the
radiation calculation reach $\sim$ 100 K.  In the latter, because the heated
cores are turbulent, a range of densities is heated to various
temperatures such that gas temperature is not a single valued function of
gas density. The heating also varies with the number of stars and
their instantaneous accretion rate, such that any EOS
fit to a mean temperature-density curve would be a generally poor
approximation over the course of the simulation. 

\begin{figure}
\vspace*{-0.5 cm}
\includegraphics[width=5.3in]{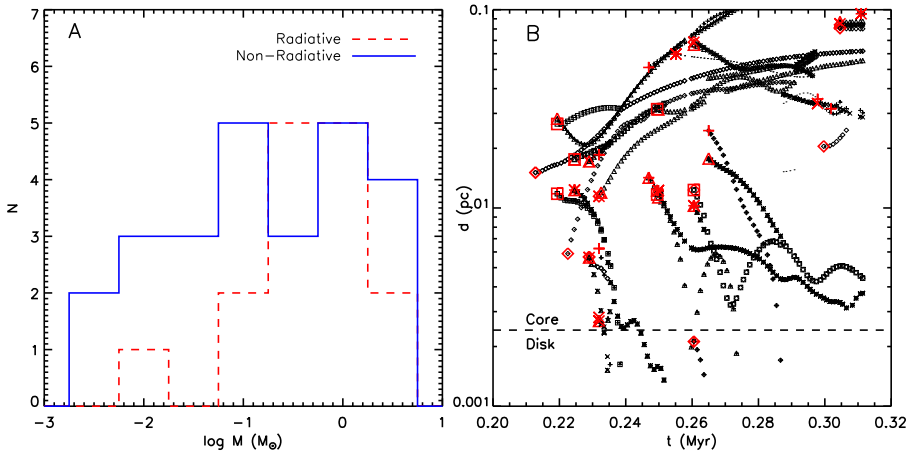}
\vspace*{-0.5 cm}
 \caption{Left: Distribution of star masses produced from
 simulations with (dashed) and without (solid) radiative feedback. Right: Pair
 separation as a function of time in 1 kyr bins for all particle pairs
 from \cite[Offner et al. (2010)]{Offner10}. The dashed line at 500 AU indicates a 
 rough boundary between the disk and core scales. The large
 majority of pairs have separations $>0.1$ pc
and are not shown. The large (red) symbols indicate the first
time bin.}
\label{imf}
\end{figure}

As a consequence of the feedback, the calculation with radiation has a
fundamentally different mass distribution (Figure
\ref{imf}A). The protostellar heating is sufficient
to raise the temperature of the gas significantly within a few hundred
AU and suppress fragmentation in accretion disks that
would otherwise be unstable. As a result, the radiative
calculation has far fewer brown dwarfs and, despite small number
statistics, its mass
distribution more closely resembles the stellar initial mass
function. 
Bate (2009) has previously demonstrated similar results using
smoothed-particle hydrodynamics simulations.    

Although the heating is efficient at small scales, it is limited
to the parent core and does not inhibit fragmentation of other
cores or even wide fragmentation within the same core. Figure \ref{imf}B
shows the separations of all star pairs as a function of time in the
calculation with radiation feedback. The plot is restricted to
separations of 0.1 pc or less to highlight stars forming
within the same core that may comprise a binary. The
plot shows that fragmentation on scales of $\sim$ 2000 AU is not
suppressed by heating. The one instance of fragmentation with
$d<500$ AU is actually filament rather than disk fragmentation. This suggests that turbulent
core fragmentation remains a viable mechanism for binary formation of
low-mass stars,
while disk fragmentation is much less likely.

\subsection{Disk Analysis}

Although the disks in the simulations are not well resolved, it is
still possible to use their mean properties to draw robust conclusions
about the protostellar multiplicity.
\cite[Kratter et al. (2010)]{kratter10} define a two-dimensional parameter space for
characterizing accretion and disk stability:
\begin{equation}
\xi = {{\mdin G}\over{\csd^3}}, ~~~~~~~~\Gamma = {{\mdin}\over{\msd\omkin}} = {{\mdin \avg{j}_{\rm in}^3}\over{G^2 \msd^3}}, 
\end{equation} 
The thermal parameter, $\xi$, compares the core sound speed to the
disk sound speed, $\csd$, where $\mdin$ is the infall mass accretion rate. 
A collapsing isothermal sphere has $\xi \simeq 1$ (\cite[Shu 1977]{shu77}). 
For $\xi >1$, a disk will be unable to 
efficiently process
accreting material and will eventually fragment. The rotational
parameter, $\Gamma$, compares 
the disk orbital time to the gas infall time, where  $\msd$ is the total mass in the star-disk system, $\omkin$ is
the Keplerian angular velocity at the circularization radius of the infall, and $\avg{j}_{\rm in}$ is the specific angular
momentum.  For large 
$\Gamma$ ($\Gamma \sim 0.1$) the disk mass changes quickly over an orbital time. 

\begin{figure}
\vspace*{-0.5 cm}
\begin{center}
 \includegraphics[width=3.0in]{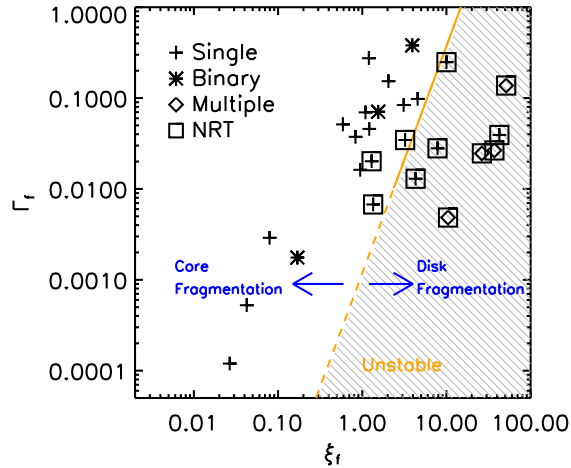}
\vspace*{-0.4 cm}
\caption{ The values of $\Gamma$ and $\xi$ at 1 freefall time for the
protostellar disks in each simulation, where the Non-Radiative
Transfer (NRT) cases are denoted by boxes. The diagonal line
indicates the boundary between stable and unstable disks found by \cite[Kratter et al. (2010)]{kratter10}.
}\label{multiples}
\end{center}
\end{figure}

In Figure \ref{multiples} we estimate these parameters for the
disks in each calculation. 
Our analysis confirms that disks in the radiative
calculation are stable, where the binaries to the left
of the line are in fact products of turbulent rather than disk fragmentation. Without radiation feedback large unstable disks yield
high multiplicity systems, which fall to the right of the line as
expected. Although some single systems exist within the unstable
regime these tend to have low Q values and often have previously fragmented. 

\section{Conclusions}

Radiation feedback from low-mass stars is important in shaping the
stellar mass distribution. Heating works to
stabilize protostellar disks and suppress fragmentation that
would otherwise over-produce brown
dwarfs. However, fragmentation of the parent core may still occur on
thousand AU scales, suggesting that turbulent core
fragmentation, not disk fragmentation, is the most likely origin of low-mass binaries.

\end{document}